\documentstyle[11pt,AATS,epsf]{article}	

\begin{document}	

\title{
Faint End of the Galaxy Luminosity Function: ~~~~~~~~~~~~~~~~~
A Chronometer for Structure Formation?
} 

\author{R. Brent Tully}
\affil{Institute for Astronomy, University of Hawaii}
\author{}
\affil{}

\begin{abstract}

There is accumulating evidence that the faint end of the galaxy luminosity 
function might be very different in different locations.  The luminosity 
function might be sharply rising in rich clusters and flat or declining
in regions of low density.  If galaxies form according to the model of
hierarchical clustering then there should be many small halos compared to
the number of big halos.  If this theory is valid then there must be a
mechanism that eliminates at least the visible component of galaxies in
low density regions.  A plausible mechanism is photoionization of the
intergalactic medium at a time before the epoch of galaxy formation in low
density regions but after the epoch of formation for systems that ultimately
end up in rich clusters.  The dynamical timescales are found to 
accommodate this hypothesis in a flat universe with $\Omega_m < 0.4$.

If this idea has validity, then upon surveying a variety of environments it 
is expected that a dichotomy will emerge.  There should be a transition
between high density / high frequency of dwarfs to lower density / low
frequency of dwarfs.  This transition should ultimately be understood by
the matching of three timing considerations: (i) the collapse timescale of the
transition density,
(ii) the timescale of reionization, and (iii) the linkage given by the 
cosmic expansion timescale as controlled by the dark matter and dark 
energy content of the universe.

\end{abstract}


\section{Introduction}

This discussion summarizes ideas developed by Tully et al. (2001).
According to the popular cold dark matter (CDM) hierarchical clustering model 
of galaxy formation
there should be numerous low mass dark halos still around today.  The 
approximation by Press \& Schechter (1974) that initial density fluctuations
would grow according to linear theory to a critical density and then
collapse and virialize leads, with a CDM-like power spectrum, to a prediction 
of sharply increasing numbers of
halos at smaller mass intervals.  Cosmological simulations are now
being realized with sufficient mass resolution to distinguish dwarf galaxies 
and this modeling basically confirms expectations of the existence of 
low mass halos (Klypin et al. 1999; Moore et al. 1999).  

Indeed, dwarf galaxies are found in abundance in some environments.  
In the past, most observational effort has gone into studies in rich clusters 
because the
statistical contrast is highest against the background (Smith, Driver, \&
Phillipps 1997;
Trentham 1998; Phillipps et al. 1998; also the small but dense Fornax Cluster:
Kambas et al. 2000).  The general conclusion from these
studies has been that, yes, there are substantial numbers of dwarfs of the
spheroidal type.  There would seem to be reasonable agreement with expectations
of CDM hierarchical clustering theory.

However, there has been a suspicion that there might not be the expected 
abundance of dwarfs in environments less extreme in density than the rich 
clusters.  Klypin et al. (1999) and Moore et al. (1999) have pointed out the 
apparent absence of
large numbers of dwarfs in the Local Group.  It is to be appreciated that
the task of identifying extreme dwarfs is not trivial.  They are tiny and
faint.  At substantial distances their surface brightnesses are faint against
the sky foreground and close up they resolve into swarms of very faint stars.
So dwarfs were not being found in the expected numbers but is this because
of observational limitations?

\section{The Ursa Major Cluster}

Motivated by the speculation that the occurrence of dwarfs might be correlated
with local density, we made extensive observations in the nearest environment
where the density is low (dynamical time is long) yet where there are enough
galaxies for a meaningful statistical discussion.  We studied the Ursa Major
Cluster, a structure fortuitously at about the same distance as the Virgo 
Cluster and which subtends a comparable amount of sky.  The total light
in bright galaxies in Ursa Major is about 1/4 that in Virgo but dynamical 
evidence suggests that the mass in Ursa Major is down by a factor 20 from 
that associated with Virgo (Tully \& Shaya 1998).  Roughly 12 sq. deg. of the 
Ursa Major Cluster 
were surveyed with deep CCD imaging with wide field cameras on the 
Canada-France-Hawaii Telescope and in the 21cm Hydrogen line with the 
Very Large Array.  The footprint of our observations is shown in Figure~1.
Results of the two aspects of the survey are 
reported respectively by Trentham, Tully, \& Verheijen (2001) and Verheijen
et al. (2000).
The important conclusion is that the luminosity function is flat at the faint
end in the Ursa Major Cluster, as seen in Figure~2.  Whereas Phillipps et al. 
(1998) found 
$\sim 700$ galaxies per sq. deg. with $-16<M_R<-11$ in Virgo, we find
$\sim 3$ galaxies per sq. deg. in the same magnitude interval in Ursa Major.
At the bright end, at $M_R<-17$, the number density of galaxies in Virgo is 
only 2.5 times higher than in Ursa Major so there is a relative difference of
two orders of magnitude in counts at the
faint end of the luminosity function between the two locations.  The VLA
survey confirms that there is no significant population of faint but HI rich
systems in Ursa Major.

\begin{figure}[htb]
\plotfiddle{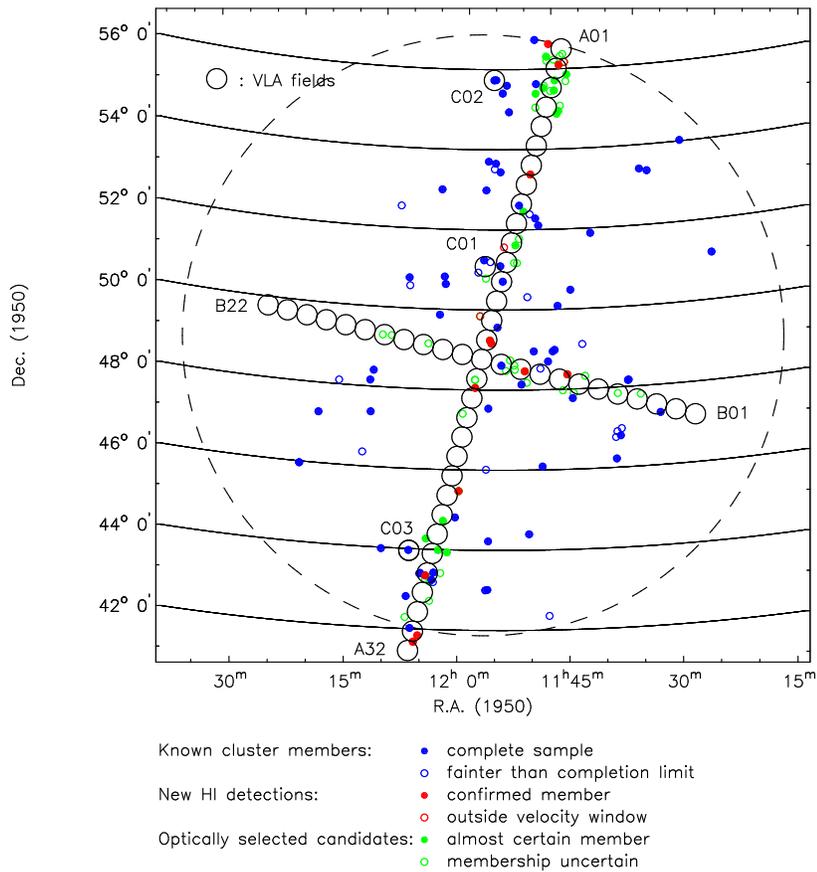}{4.2in}{0}{59}{59}{-180}{-50}      
\caption{Ursa Major Cluster survey.  The area covered by the VLA survey and
with comparable fields by the CFHT wide field CCD survey is indicated by the
pattern of circles.  Before the survey began, 79 galaxies were known to be
associated with the Ursa Major Cluster.  The VLA HI survey detected only 10
more members within the survey footprint.  The CCD survey has revealed another
3 dozen probable or possible members.
\label{1} }
\end{figure}

\begin{figure}[htb]
\plotfiddle{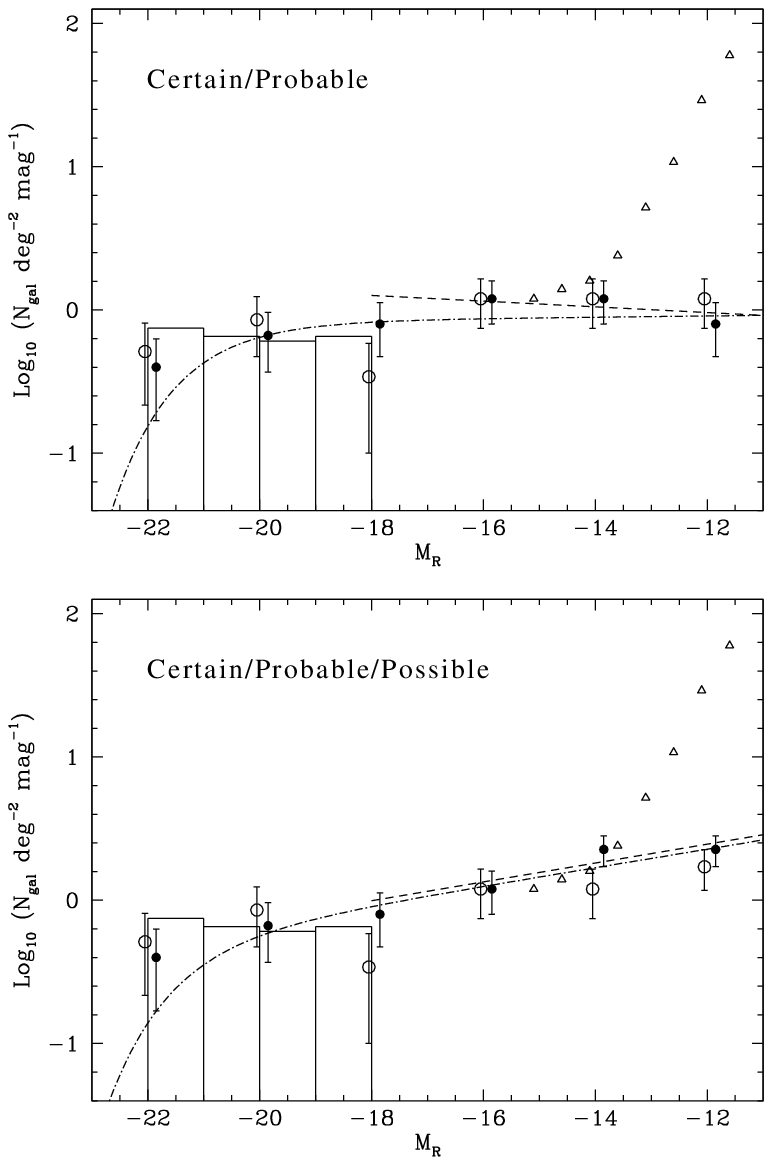}{4.2in}{0}{100}{100}{-290}{-240}      
\caption{Ursa Major Cluster luminosity function.  Histogram is the luminosity
function for the complete bright sample and the points with error bars pertain
to the area of the cluster covered by the VLA/CCD survey.  The top panel
only includes certain and probable cluster members while the bottom panel
also includes possible cluster members.  Small triangles illustrate the 
flair-up at faint magnitudes found in the Virgo Cluster by Phillipps et al.
\label{2} }
\end{figure}

The Ursa Major luminosity function resembles the luminosity function of the
Local Group and, indeed, of other nearby groups.  Normalizing to the 
occurrence of luminous galaxies, there is a shortfall of one to two 
orders of magnitude from the numbers of dwarfs seen in the Virgo
Cluster.  The Ursa Major Cluster has a lot of galaxies but in other 
respects it resembles the nearby groups.  It is a loose irregular cluster
filled with HI rich spirals with a crossing time comparable to a Hubble time.
From the evidence at hand, such environments host relatively
few faint dwarfs.  Yet other environments with short crossing times, like
Virgo, Fornax, and rich Abell clusters, seem to have large numbers of
dwarfs.

\section{Squelched Galaxies}

Hierarchical clustering theory anticipates that there should be numerous
dwarf galaxies relative to giant galaxies and this situation is found in 
rich clusters.   This theory predicts that the relative number of dwarfs is
even higher in low density regions (Sigad et al. 2000) yet far fewer are
found.
Apparently we need to explain the {\it absence} of small
galaxies in low density environments.  At first thought, it would seem that
the rich clusters are more hostile, the low density regions more benign for
the survival of small galaxies.  In very low density groups dynamical 
collapse times can be of order the age of the universe and many galaxies
should not have had time to interact with any other galaxy.  Hence probably
the answer to our problem does not lie with tidal interactions between
systems.  We need to call upon a mechanism that {\it allows} small galaxies 
to form in rich clusters but {\it thwarts} small galaxy formation in places 
of low density.

A plausible squelching mechanism is 
photoionization of the intergalactic medium before the epoch of galaxy
formation.  Efstathiou (1992) discussed the inhibiting
effect on the formation of dwarfs due to the suppression of cooling of a 
primordial plasma of hydrogen and helium.  Thoul \& Weinberg (1996) took the 
discussion further with recourse to high resolution hydrodynamic simulations.
These authors argue that gas heating before collapse is more important than
inhibition of line cooling.  The suppression of galaxy formation occurs
below a mass threshold.  The UV background heats the precollapse gas to
roughly 25,000~K.  This temperature is much less than that associated with
the virial energy of a large galaxy, hence has negligible effect on the 
collapse of baryons into a massive potential well.  However, for a sufficiently
small galaxy this heating is comparable with, or can dominate, the 
gravitational energy.  Thoul \& Weinberg find there is essentially total 
suppression
of baryon collapse for systems with circular velocities 
$V_{circ}<30$~km~s$^{-1}$ and, by contrast, little effect on galaxy 
formation for systems with $V_{circ}>75$~km~s$^{-1}$.  It follows that
luminosity functions would be unaffected above 
$M_R^{b,i}\sim-18.6+5{\rm log}h_{75}$
($M_B^{b,i}\sim-17.8+5{\rm log}h_{75}$) but truncated below 
$M_R^{b,i}\sim-16$ ($M_B^{b,i}\sim-15$). Here, $h_{75}={\rm H}_{\circ}/75$
and superscripts $b,i$ indicate corrections are made for Galactic and internal
obscuration.

The Thoul \& Weinberg model assumes galaxy collapse occurs after reheating of 
the intergalactic medium.  The collapse timescale (Gunn \& Gott 1972) is
$$
t_{col}=1.4\times10^{10}(R_{vir}^3/M_{14})^{1/2} h_{75}^{-1}~{\rm yr}
$$
where $R_{vir}$ is the virial radius in Mpc and $M_{14}$ is the virial mass in 
units of $10^{14}~M_{\odot}$.  
Values for $R_{vir}$ and $M_{14}$ can be extracted from
Tully (1987) for the Virgo and Ursa Major clusters ($R_{vir}$: 0.79 and 
0.98~Mpc
respectively; $M_{14}$: 8.9 and 0.5 respectively).  Hence, rough dynamical
collapse times for these clusters are $t_{col}^{virgo}\sim3.3$~Gyr and
$t_{col}^{uma}\sim19$~Gyr.   The dense, elliptical dominated Virgo Cluster 
formed a {\it core} 
long ago and the loose, spiral dominated Ursa Major Cluster is still in the
process of collapsing.  Of course, galaxies continue to fall in and enlarge
the Virgo Cluster to this day and, on the other hand, 
substructure in Ursa Major would have shorter dynamical collapse times than
the entire entity.

Smaller mass scales collapse before larger mass scales.  Dwarfs must form
before their host cluster form.  For the present discussion, the rough 
approximation is assumed that structure on the mass scale of dwarfs formed 
at $\sim 1/3$ the time of the collapse of the cluster core.  The progression
of hierarchical collapse and merging can be followed in semi-analytic models
(eg, Somerville \& Primack 1999; Springel et al. 2000).  Elaborations on these
points will
be provided in the discussion by Tully et al. (2001).

These formation timescales in hand, we now
ask whether the dwarfs should have formed before or after reionization of the
intergalactic medium by the UV radiation of AGNs or hot stars.  Observations
constrain the epoch of reionization to $z>6$ (Fan et al. 2000), which can be
understood on theoretical grounds (Gnedin \& Ostriker 1997).
In Figure~3 we see the relationship between
redshift and the age of the universe for a wide range of topologically flat
cosmological models.  If baryon collapse into small galaxies can only occur
before reionization then Fig.~3 tells us that if the epoch of reionization 
is as late as $z_{ion}\sim6$ then dwarfs with $t_{col}\sim1$~Gyr could 
form in a universe with 
matter density $\Omega_m\sim0.2$ and vacuum energy density 
$\Omega_{\Lambda}\sim0.8$.

\begin{figure}[ht!]
\plotfiddle{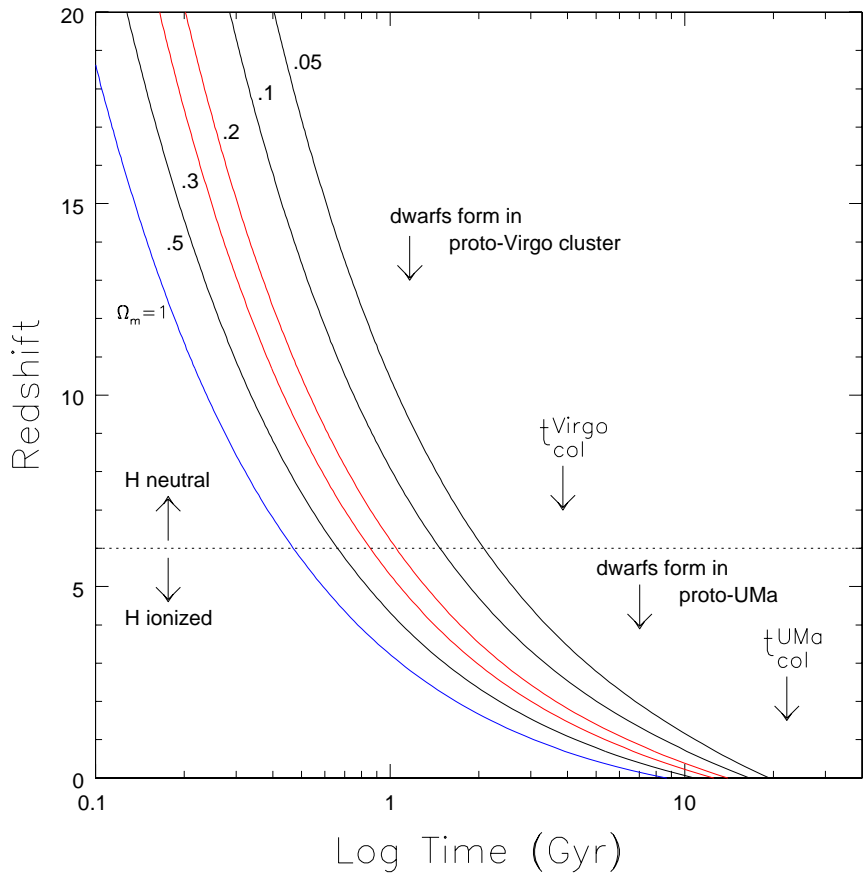}{4.2in}{0}{60}{60}{-180}{-93}
\caption{Redshift vs age of the universe for a range of flat world models,
from $\Omega_m=1$, $\Omega_{\Lambda}=0$ on the bottom to 
$\Omega_m=0.05$, $\Omega_{\Lambda}=0.95$ on top.  The arrows
indicate the rough epochs of galaxy formation in the Virgo and Ursa Major
clusters and the collapse timescales of the clusters.  Intergalactic 
reionization must have occurred at $z_{ion}>6$, that is, above the 
horizontal dotted line.}
\end{figure}

We conclude that it is very plausible that small mass halos in the proto-Virgo
region collapsed before reionization but almost certainly small mass halos in 
the proto-Ursa Major region collapsed after the universe was reionized.
Hence this single mechanism could explain why there are many visible dwarf 
galaxies in dense environments and few in low density regions.
Interestingly, this mechanism only works in a universe with relatively low
matter density, say $\Omega_m<0.4$, $\Omega_{\Lambda}>0.6$.  In a universe
with $\Omega_m=1$, structure forms at low redshift: $t_{col}\sim1$~Gyr 
corresponds to $z\sim3$.

It would follow that if a range of cluster environments is explored then
there should be a break: denser clusters with short dynamical times will
have many dwarfs and less dense clusters with long dynamical times will have
few dwarfs.  The collapse time scale associated with the break point density
would reflect the time of reionization of the universe.

\section{Summary}

\noindent
1. The faint end of the luminosity function of galaxies might be steeply
rising in the dense environment of rich clusters but flat or falling in
the low density regions of groups.  Galaxy formation models anticipate
the mass function is sharply rising at the low mass end.  It seems something
is suppressing the visible manifestations of small galaxies in low density 
environments.

\noindent
2. Reionization of the universe at $z_{ion}>6$ could inhibit the collapse of 
gas in low mass potential wells for late forming galaxies.  Dynamical collapse
times inferred from the observed densities of clusters are consistent with
the picture that dwarf halos formed {\it before} reionization in high density 
regions and {\it after} reionization in low density regions, but only if 
structure is forming at high redshift; ie, $\Omega_m < 0.4$ in a flat
universe.




\noindent
\acknowledgments

My collaborators in this research are Rachel Somerville, Neil Trentham,
and Marc Verheijen.  Financial 
support has been provided by a NATO travel grant.

\end{document}